# Laser Induced Periodic Surface Crystalline Patterns on SrO-0.5Li$_2$O-4.5B$_2$O$_3$ and BaO-0.5Na$_2$O-4.5B$_2$O$_3$ Glasses and Optical Second Harmonic Generation

*Rahul Vaish[a], Vincent Rodriguez[b], Mario Maglione[c], J. Etourneau[c], K.B.R. Varma[a]\**

[a]Materials Research Centre, Indian Institute of Science, Bangalore, India

[b]Laboratoire de Physico—Chimie Moleculaire, UMR 5803 CNRS—Universite, Bordeaux, France

[c] Institut de Chimie de la matière condensée de Bordeaux-CNRS, France

*Corresponding Author; E-Mail : kbrvarma@mrc.iisc.ernet.in;

FAX: 91-80-23600683; Tel. No: 91-80-22932914

**Abstract:**

Transparent $SrO$-$0.5Li_2O$-$4.5B_2O_3$ (SLBO) and $BaO$-$0.5Na_2O$-$4.5B_2O_3$ (BNBO) glasses were fabricated via the conventional melt-quenching technique. The amorphous and glassy characteristics of the as-quenched samples were respectively confirmed by X-ray powder diffraction (XRD) and Differential Thermal Analysis (DTA) studies. Optically clear SLBO and BNBO glass nano/microcrystal composites were obtained by heat-treating the as-quenched glasses slightly above their respective glass transition temperatures. These glass nano/microcrystal composites exhibited second harmonic signals in transmission mode when exposed to infrared laser beam at $\lambda$=1064 nm. Interestingly second-harmonic waves were found to undergo optical diffraction and was attributed to the presence of self-organized microcrystalline patterns associated with 0.5-2µm periodicities.

## I. Introduction

Glasses of a variety of nonlinear/ferroelectric materials are being widely used for multifarious applications as these could be fabricated with ease in large sizes with high optical homogeneity accompanied by high degree of transparency over wide range of wavelengths of light. It would be interesting from the view point of practical applications to obtain glasses associated with promising physical properties. In the past several years, lot of efforts have been made to bring in 3-D microscopic modification in transparent materials using lasers [1, 2]. The promising applications include 3-D optical memories, optical waveguides, couplers, laser beam induced nanogratings etc. The interaction between the light and glass is generally resulted in various types of light induced structural changes and crystallization. Nano/micro gratings with various structures are indispensable for optical diffractive elements and non-linear optical devices. Ultrafast laser could be used for fabricating nano/micro patterns on a variety of materials for ever-increasing applications especially as optical and opto-electronic elements.

Borate-based crystals were proved to be the key materials for photonic applications. However, efforts are being made to visualize the usefulness of their glass counter parts for the same applications. Borate-based materials in general have drawn the attention of many researchers because of their large non-linear optical coefficients, wide transmission range, high laser induced damage threshold, low material cost, good chemical and mechanical stability and moderate melting points [3]. Recently, the details concerning the structures of single crystals of $SrLiB_9O_{15}$ and $BaNaB_9O_{15}$ have been reported [4]. These were found to crystallize in non-centrosymmetric space group R3c. The compositions for the present study are chosen in such a way that one would eventually obtain $SrLiB_9O_{15}$ and $BaNaB_9O_{15}$ crystalline phases on crystallization of $SrO$-$0.5Li_2O$-$4.5B_2O_3$ (SLBO) and $BaO$-$0.5Na_2O$-$4.5B_2O_3$ (BNBO) glasses, respectively. It would be interesting to introduce microcrystallinity in these borate glasses to visualize their non-linear optical effects which include second harmonic generation and Pockels effect. In this article, the fabrication of self-organized periodic microcrystalline patterns has been demonstrated using infrared (1064 nm) laser radiation that is weakly absorbed by the SLBO and BNBO glasses.

**II Experimental**

Transparent SLBO and BNBO glasses were fabricated from their constituent carbonates/oxides. For this, the appropriate amounts of carbonates ($Li_2CO_3$, $Na_2CO_3$, $SrCO_3$ and $BaCO_3$) and $H_3BO_3$ were thoroughly mixed and melted in a platinum crucible in the 1273-1373K temperature range depending upon the composition. Melts were quenched by pouring on a steel plate that were preheated to 423K and pressed with another one to obtain 1-2mm thick transparent plates. All these samples were annealed at temperatures which are about 50K below their respective glass transition temperatures for 5 h to minimize the residual stresses. The glassy nature of the as-quenched samples was confirmed by subjecting the samples to non-isothermal differential thermal analysis (TA Instruments) at a heating rate of 10 K/min. X-ray powder diffraction (Philips PW1050/37) studies were carried out on the as-quenched samples using Cu-K$a$ radiation. The optical transmission spectra of these samples were recorded using a spectrophotometer (Bruker IFS 66v/s Vacuum Fourier Transform Interferometer) in the 100–1200 nm wavelength range. The second harmonic (532 nm) intensities from the heat-treated samples were recorded at 300 K using the Maker-Fringe method. Q-switched Nd:YAG laser operating at 1064 nm wavelength was used as the fundamental light source. The laser irradiated samples were examined using optical microscope (Olympus BX51) in transmission mode.

**III Results and discussion**

The DTA curves that were obtained for the as-quenched bulk glasses are shown in Fig. 1 ((a) and (b)). Glass transition temperatures ($T_g$) for SLBO and BNBO glasses are 816K and 795K respectively and their crystallization temperatures ($T_{cr}$) are 930K (SLBO) and 890K (BNBO) respectively. The X-ray powder diffraction (XRD) patterns obtained for the as-quenched glasses are depicted in Fig 2((a) and (c)) which confirm their amorphous nature. The XRD patterns recorded for the SLBO and BNBO glasses heat-treated (848 K/5 h and 813 K/10 h, respectively) just above their respective glass-transition temperatures are also depicted in Fig. 2((b) and (d)). The Bragg peaks in these XRD patterns (Fig. 2 (b) and (d)) are found to correspond to the $SrLiB_9O_{15}$ and $BaNaB_9O_{15}$ crystalline phases, respectively. An attempt has been made to estimate the crystallite size based on these XRD studies using Williamson-Hall method [5]. The crystallite sizes that were obtained for heat-treated SLBO and BNBO samples were around 10 nm and 80 nm respectively.

The optical transmission spectra (uncorrected for reflection losses) were recorded for all the as-quenched and heat-treated samples (comprising nanocrystallites) in the 100-1200 nm wavelength range. Fig. 3 ((a) and (b)) shows the transmission spectra for all the samples (as-quenched and heat-treated) under study. In order to have further insight into the optical properties of these glasses and glass-nanocrystal composites, the optical band gap was calculated using the optical transmission data. The optical absorption coefficient ($\alpha$) is related to the intensity of light transmitted out ($I$) of a sample of thickness, $t$, as:

$$\alpha = \frac{1}{t} \ln\left(\frac{I_o}{I}\right) \quad (1)$$

where $I_o$ is the intensity of the incident light on the sample. The study of optical absorption may be used as a probe to study the electronic structures of the glasses. In the high absorption region, Tauc [6] and Davis & Mott [7] suggested an equation for the absorption coefficient ($\alpha$) and the incident photon energy ($h\nu$) as:

$$\alpha = \frac{B(h\nu - E_{opt})^n}{h\nu} \quad (2)$$

where $B$ is the constant, $E_{opt}$ is the optical band gap energy and exponent $n$ could be taken as 2 and ½ for an indirect and direct optical transitions, respectively. In the present study, $(\alpha h\nu)^{1/2}$ was plotted against ($h\nu$) (Fig. 4) for the as-quenched and heat-treated glasses. It is clear from the figure that the plots of $(\alpha h\nu)^{1/2}$ versus ($h\nu$) have linear segment (in high absorption region) for all the samples under investigation. The value of $E_{opt}$, was determined for all the fabricated glasses and glass nanocrystal composites using the above equation (Eq.2) for n = 2 by extrapolating the linear fitting of the high absorption regions to zero absorption (($\alpha h\nu)^{1/2}$=0) as shown in the same figure (Fig. 4) with solid lines. The values obtained for $E_{opt}$ for SLBO as-quenched and heat-treated glasses are 3.4 eV and 3 eV whereas 3.1 eV and 2.8 eV were obtained for BNBO as-quenched and heat-treated glasses, respectively. Absorption is observed in amorphous materials below the band gap because of the transitions from occupied extended states of the valence band to empty tail states of the conduction band and from occupied valence band tail states to the empty extended states of the conduction band. The width of these band tails extended into the band gap is expressed in terms of Urbach energy and is given as [8]:

$$\alpha = \alpha_o \exp(h\nu/\Delta E) \tag{3}$$

where $a_o$ is a constant and $?E$ is the Urbach energy which indicates the width of the band tails of the localized states and associated with the structural disorder of the glasses. The plots of ln $a$ versus photon energy ($h?$) for all the glasses are shown in Fig. 5. The values obtained for the Urbach energy are calculated from the slopes of the linear portions of these curves. The values for $?E$ for SLBO as-quenched and heat-treated glasses are 0.6 eV and 0.65 eV whereas 0.82 eV and 1.05 eV were obtained for the as-quenched and heat-treated BNBO glasses, respectively. The glasses with larger Urbach energy would have greater tendency to convert weak bonds into defects. Consequently, the defect concentration could be decided by the measure of Urbach energy. Among the SLBO and BNBO glasses, the least Urbach energy is obtained for the SLBO glasses. This suggests that the defects in these glasses are minimum and facilitating local long range order resulting in the least Urbach energy.

The heat-treated (in the vicinity of $T_g$) SLBO (comprising 10 nm sized crystallites) and BNBO (comprising 80 nm sized crystallites) glasses were exposed to Nd:YAG laser (1064 nm) and second harmonic signals (SHG) (transmission mode, pp polarization) were recorded as a function of the angle of incidence. The SHG intensity collected as a function of the incident angle for the heat-treated SLBO and BNBO glasses are shown in Fig. 6 and Fig. 7 respectively. The second harmonic signal was found to undergo an optical diffraction. In order to understand the underlying mechanism of the optical diffraction of the second harmonic signals, the samples were examined under optical microscope subsequent to SHG experiments. Interestingly, the presence of fine parallel lines running across the sample was noticed on the surfaces and found to penetrate deep into the samples depending on the laser exposure time and its power. Optical micrographs of the laser induced periodic structures are also shown in Fig. 6 and Fig. 7 for the heat-treated SLBO and BNBO glasses respectively.

An attempt has been made to rationalize the above results. The nanocrystallites associated with the heat-treated glasses, act as light scattering centers. When the laser beam propagates with a component velocity along the surface of the sample it gets scattered. The interference between the scattered and incident radiation that occurs along the axis of the scatterers leads to the formation of the interference fringes observed only at certain angles which correspond to the axis of these scatterers, when the sample is tilted as a function of the incident angle [9-11]. In other words, the damage fringes that are produced parallel to the scatterers in the present samples have been attributed to the interference of the incident beam with the surface-scattered waves originating from the scattering centers [12, 13]. The parallel damage fringes that are encountered on the glass under study are separated by a distance (*d*) equals to [14]

$$d = \frac{\lambda}{1 \pm \sin\theta} \qquad (4)$$

where $\lambda$ is the wavelength of the incident light and $\theta$ is the angle of incidence measured from the surface normal. The interfringe spacings that are found experimentally from micrographs (Fig.6 and Fig. 7) are in good agreement with those predicted by the above formula (dotted lines in the above figures). Closer examination of the surface revealed that these fringes are actually consisting of a row of equally spaced fine crystallites of submicrometer (0.2 to 0.8 $\mu$m) size, though the samples prior to laser exposure had only nanometer sized crystallites. The increase in crystallite size from nano to micrometer level in the irradiated samples is ascribed to the localized heating effects caused by the intense input laser radiation. The presence of this submicrometer sized crystal accounts for the diffraction effects that are encountered.

The rise in temperature is estimated on the surface of the samples by using the following formula [15]

$$\Delta T = \frac{W(1-R)\alpha}{C} \qquad (5)$$

where $W$ (38 J/cm$^2$) is the laser fluence. $C$ the volume specific heat. The C values that are obtained for SLBO and BNBO glasses respectively are 2.55 J/cm$^3$.K and 2.81 J/cm$^3$.K. $R$ is the reflection coefficient and a the absorption coefficient.

The absorption coefficient (a) is calculated using the equation [16]

$$\alpha = \frac{1}{t}\ln\left[\frac{(1-R)^2}{2T} + \sqrt{\frac{(1-R)^2}{2T} + R^2}\right] \qquad (6)$$

where $t$ is the thickness of the sample, $T$ is the transmisstance and $R$ is the reflectance which is evaluated based on the refractive index data. For normal incidence the reflectance ($R$) could be described by the following formula

$$R = \frac{(n-1)^2 + k^2}{(n+1)^2 + k^2} \qquad (7)$$

where n is the refractive index of the glass and k is the extinction coefficient which could be neglected for transparent glasses. For the SLBO and BNBO glass nanocrystal composites, the measured values of $n$ (at 1064 nm) are 1.529 and 1.504, respectively. After neglecting k, the above equation becomes

$$R = \frac{(n-1)^2}{(n+1)^2} \tag{8}$$

The value of reflectance for SLBO and BNBO glass-nano/microcrystal composites is found to be around 0.045 and 0.04, respectively. The absorption coefficient was calculated using Eq.6. and found to be 0.07/$t$ and 0.141/$t$ for SLBO and BNBO glass-nanocrystal composites, respectively.

Using the above data, Eq. 5 could be written as

$$\Delta T = \frac{9960}{t(\mu m)} \quad \text{(For SLBO glass nano/microcrystal composites)} \tag{9}$$

and for BNBO glass nano/microcrystal composites

$$\Delta T = \frac{18360}{t(\mu m)} \tag{10}$$

It is to be noted that the temperature rise varies as a function of the depth from the surface to the interior of the sample. Fig.8 shows the variation of temperature as a function of the depth (in µm) of the sample from the surface (as evaluated from Eqs. 9 and 10). It is clear from the figure (Fig. 8) that the temperature decreases with increase in penetration depth for both the glass compositions. It is to be generally noted that the total crystallization of the glasses is possible only when

?T = $T_{cr}$  (Crystallization temperature) (11)

The crystallization temperatures are 930 K and 890 K for SLBO and BNBO glasses, respectively (as determined by DTA studies). It is evident from Fig.8 that these temperatures facilitate crystallization upto about 11 µm and 20 µm from the surfaces of SLBO and BNBO glass nanocrystal composites, respectively. It should be pointed out that though the temperatures achieved by laser irradiation are close to that of bulk crystallization temperatures of respective glasses, total crystallization did not occur as the incident-laser beam is of nanosecond duration.

The strong electric field (vector) that is associated with the incident laser beam assisted by a rise in temperature would result in aligning the crystallites. The strength of the electric field, is determined using the eq. [17]

$$E = \sqrt{\frac{2\eta_o cW}{n}} \tag{12}$$

where $E$ is the electric field vector associated with the laser, c speed of light, $W$ is laser fluence and $\eta_o$ $\left(\sqrt{\mu_o/\varepsilon_o} = 377\Omega\right)$ is the impedance of the free space. The calculated value of electric field associated with the laser beam is around $2 \times 10^7$ V/cm. This strong electric field vector facilitates the alignment of the crystallites in a subtle way on the glass surfaces as there is already an intense interference fringe along which the crystallization would take place. The formation of self-organized patterns would diffract light as the crystallite sizes (0.2 -0.8 μm) are in the same order as that of the wavelength (0.532 μm) of the SHG light propagating through the samples.

**IV Conclusions**

The optical properties of transparent SLBO and BNBO glasses and glass-nano/microcrystal composites, which are of technological importance, have been studied. Optical gratings were inscribed on SLBO and BNBO glasses using 1064 nm laser radiation. These exhibited intense second harmonic signals which were found to undergo optical diffraction. The diffraction of second harmonic intensity is attributed to the presence of well-aligned submicrometer sized crystallites. This process may be exploited for the fabrication of nano/micrograting which may find applications in nano photonic devices.

## Acknowledgments

One of the authors (R.V.) thanks the French Embassy, New Delhi for providing fellowship to carry out some of the present investigations at LPCM, Bordeaux, France.

**Figure captions**

Fig. 1: DTA traces for the as-quenched (a) SLBO and (b) BNBO glasses

Fig. 2: X-ray powder diffraction patterns for the pulverized (a) as-quenched SLBO, (b) heat-treated SLBO, (c) as-quenched BNBO and (d) heat-treated BNBO glasses

Fig. 3: Optical transmission spectra for both the as-quenched and heat-treated (a) SLBO and (b) BNBO glasses

Fig. 4: The plots of $(ah\nu)^{1/2}$ versus $h\nu$ for (a) SLBO and (b) BNBO glasses

Fig. 5: Urbach plots for the as-quenched and heat-treated SLBO and BNBO glasses

Fig. 6: Optical micrograph and the corresponding second harmonic intensities (532 nm) for the heat-treated SLBO glasses (dotted lines are the fitted values)

Fig.7: Optical micrographs and the corresponding second harmonic intensities (532 nm) for the heat-treated BNBO glasses at two different locations on the surfaces (dotted lines are the fitted values)

Fig. 8: Temperature variation in SLBO and BNBO glass nano/microcrystal composites due to laser irradiation

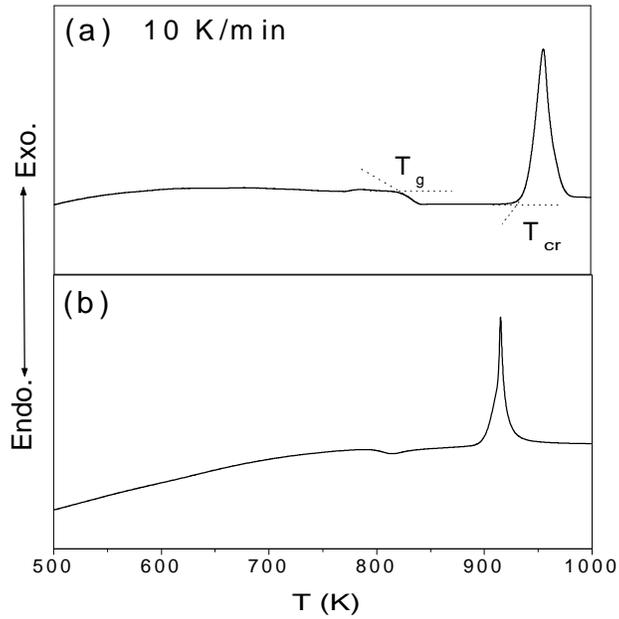

Fig. 1: DTA traces for the as-quenched (a) SLBO and (b) BNBO glasses

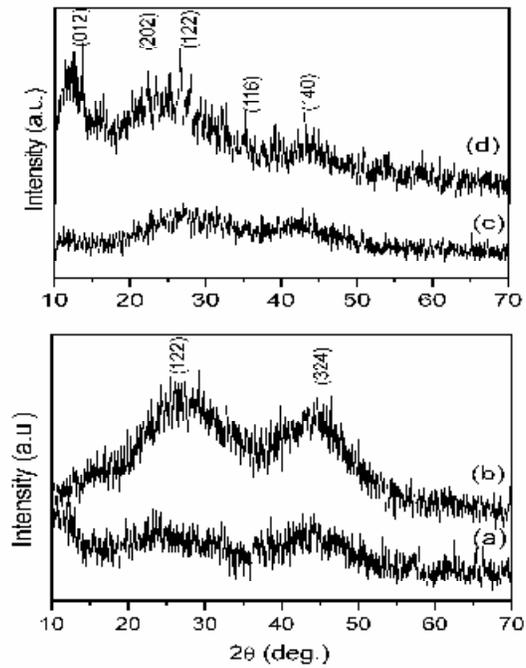

Fig. 2: X-ray powder diffraction patterns for the pulverized (a) as-quenched SLBO, (b) heat-treated SLBO, (c) as-quenched BNBO and (d) heat-treated BNBO glasses

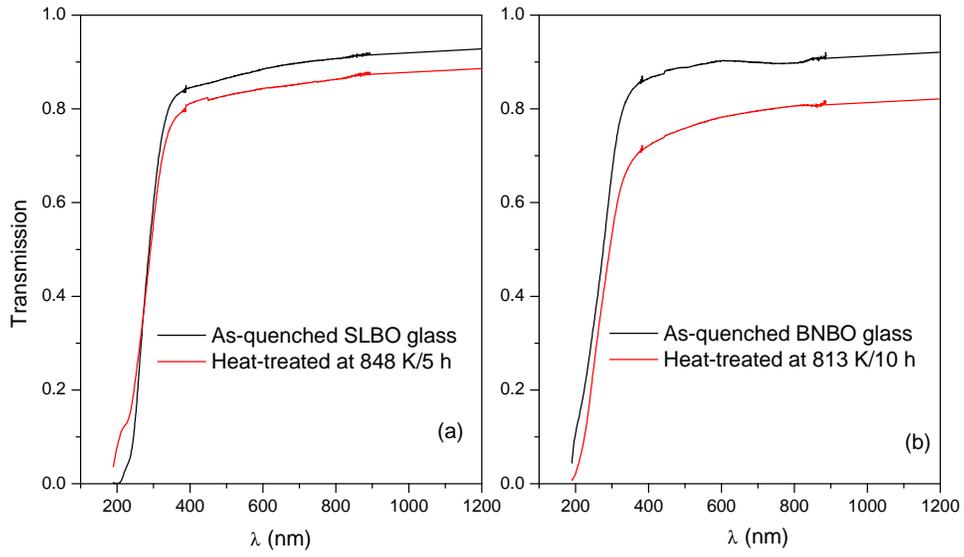

Fig. 3: Optical transmission spectra of the (a) SLBO and (b) BNBO glasses

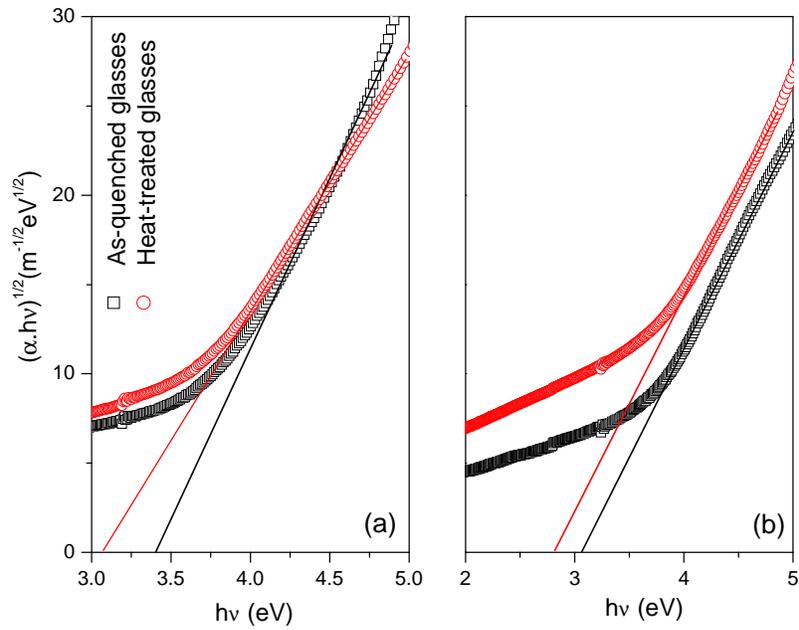

Fig. 4: The plots of $(\alpha h\nu)^{1/2}$ versus $h\nu$ for (a) SLBO and (b) BNBO glasses

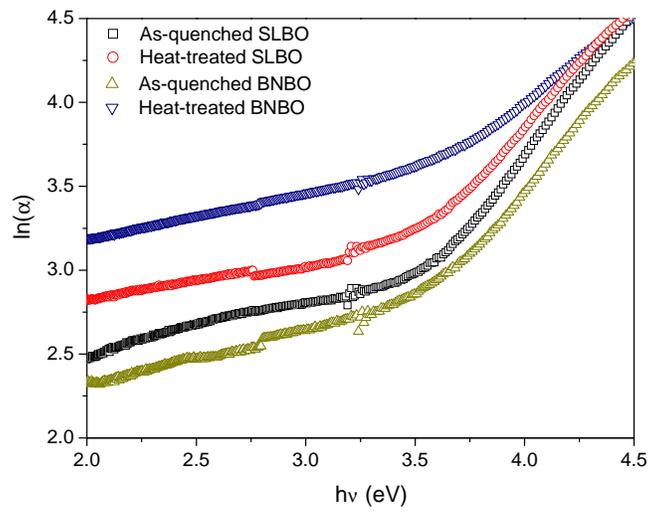

Fig. 5: Urbach plots for the as-quenched and heat-treated SLBO and BNBO glasses

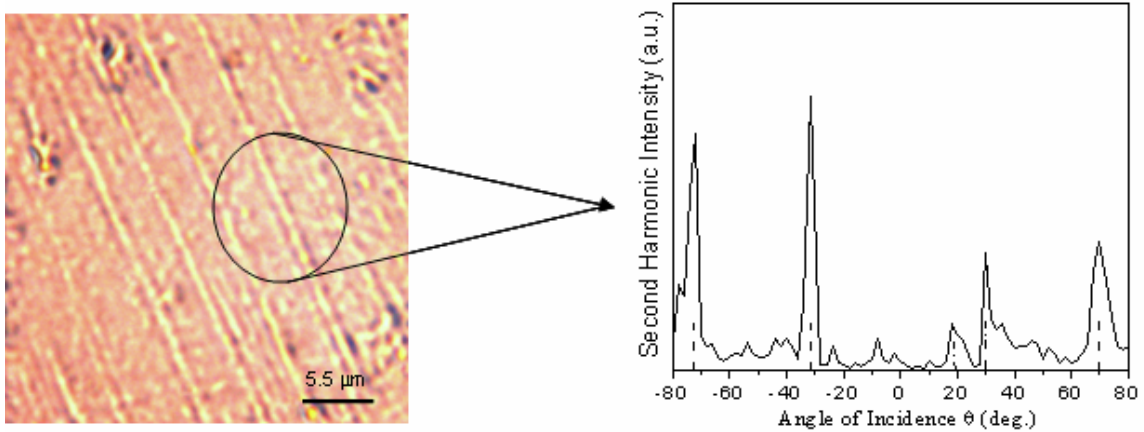

Fig. 6: Optical micrograph and corresponding second harmonic intensities (532 nm) for the heat-treated SLBO glasses (dotted lines are the fitted values)

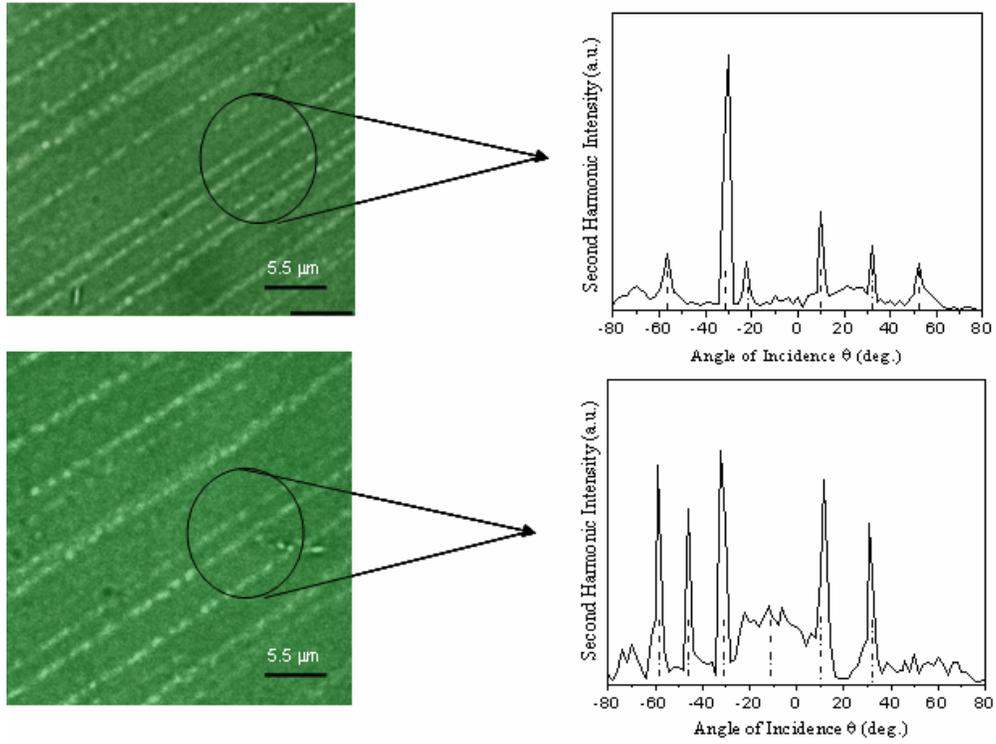

Fig.7: Optical micrographs and the corresponding second harmonic intensities (532 nm) for the heat-treated BNBO glasses at two different locations on surfaces (dotted lines are the fitted values)

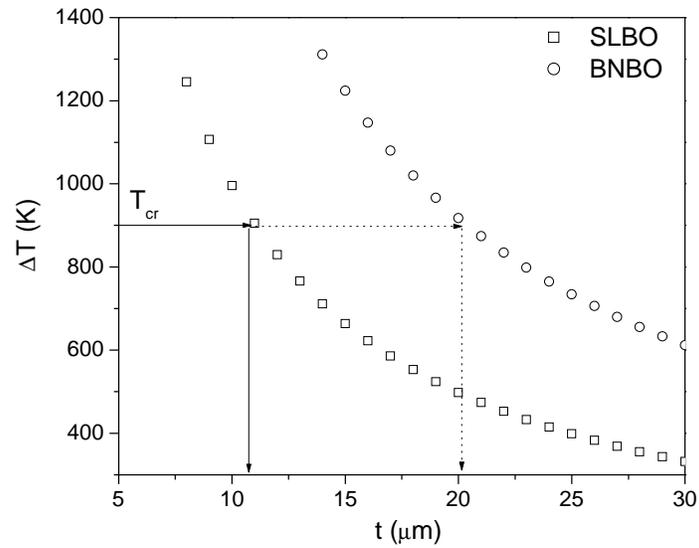

Fig. 8: Temperature variation in SLBO and BNBO glass-microcrystal composites due to laser irradiation